# Direct evidences of pentagonal silicon chains and magic clusters

Shaoxiang Sheng[1,4][†], Runze Ma[2][†], Jiangbin Wu[3,4][†], Wenbin Li[1,4], Longjuan Kong[1,4], Xin Cong[3,4], Duanyun Cao[2], Wenqi Hu[1,4], Jun-Wei Luo[3,4], Peng Cheng[1,4], Ping-Heng Tan[3,4]*, Ying Jiang[2,5]* Lan Chen[1,4] and Kehui Wu[1,4,5]*

**Affiliations:**

[1]Institute of Physics, Chinese Academy of Sciences, Beijing 100190, China

[2]International Center for Quantum Materials, School of Physics, Peking University, Beijing 100871, China

[3] State Key Laboratory of Superlattices and Microstructures, Institute of Semiconductors, Chinese Academy of Sciences, Beijing 100083, China

[4]School of physics, and College of Materials Science and Opto-electronic Technology, University of Chinese Academy of Sciences, Beijing 100049, China

[5]Collaborative Innovation Center of Quantum Matter, Beijing 100871, China

*Correspondence to: phtan@semi.ac.cn, yjiang@pku.edu.cn, khwu@iphy.ac.cn

[†]These authors contributed equally to this work.

**Pentagon is one of the most beautiful geometric structures in nature, but it is rarely seen simply because five-fold symmetry is mathematically forbidden in a 2D or 3D periodic lattices. Fortunately, pentagon as a structural element is allowed in 1D or 0D systems, since translational symmetry is not necessary there. For examples, pentagons can often be found in low-dimensional carbon systems, such as the boundaries of graphene[1-3], defects in carbon nanotube[4], or in fullerene molecule[5,6]. However, in these systems pentagons only compose a small portion of the structure. So far, 1D or 0D systems consisting of purely pentagons are still rare. Here, combing high resolution non-contact atomic force microscopy and tip-enhanced Raman spectroscopy, we have directly visualized the**

**pentagon-ring structure in self-assembled Si nanoribbons and magic clusters on Ag(110) substrate. Moreover, chemical fingerprint of Si pentagon was detected in individual Si nanoribbon and clusters by tip-enhanced Raman spectroscopy. This work demonstrates that Si pentagon can be an important element in building silicon nanostructures, which may find potential applications in future nanoelectronics based on silicon.**

It has been known that on anisotropic Ag(110) surface, silicon atoms self-assemble into 1D nanoribbons (SiNRs) aligned along the [$\bar{1}$10] direction of Ag(110) substrate, as was first reported by C. Leandri et al[7]. While at room temperature the ribbon consists of a single-strand (SNRs, 0.8 nm in width), as shown in Fig. 1a, upon annealing at about 460 K most ribbons are converted to double-strand ones (DNRs, 1.6 nm in width, Fig. 1f). It has been more than a decade since the discovery of SiNRs and more than ten structure models have been proposed to account for the atomic structures of the SiNRs[8-17], all of which were based on the scanning tunneling microscopy (STM) images combined with density functional theory (DFT) calculations. Most of these models consist of Si hexagons, and some even include mixed Ag-Si alloy structures[15]. Remarkably, a recent structural model by Cerdá et al proposed that the SNRs are composed of purely silicon pentagons[18], which was supported by a following experiment based on grazing incidence x-ray diffraction (GIXD)[19]. These two works cast new lights in resolving the intriguing SiNR structure. However, since STM reflects the electronic states instead of the exact atomic geometry, and GIXD is also an indirect probe, these two works still largely relied on theoretical modeling and cannot completely exclude possibilities other than the pentagon model, and cannot exclude the alloy model as well. A direct experimental approach is highly desired in order to pin down the pentagon model of SiNRs.

In this study, we provide two direct experimental evidences that the SiNRs on Ag(110) is composed on purely Si pentagons. Firstly, by using non-contact atomic force microscopy (nc-AFM) with the tip decorated by a CO molecule, we were able to visualize the Si pentagon rings directly. Secondly, by using an STM tip-enhanced Raman spectroscopy (TERS), we were able to resolve the local chemical fingerprint of SiNRs, which fits perfectly to Si pentagons rather than hexagons. These two experiments unambiguously pin-down the pentagonal nature of the SiNRs. Moreover, we have achieved single-cluster TERS spectrum, from Si magic clusters that co-exist with SiNRs on Ag(110) at low Si coverage. In contrast to the so-far proposed hexagonal model for Si magic clusters[18], our results based on both nc-AFM and TERS revealed that they are composed of a pair of Si pentamers.

**Results and discussion**

Nc-AFM is a powerful tool for visualizing surface structures with ultrahigh atomic resolution. The key lies in probing short-range Pauli repulsive forces, which is very sensitive to the true positions of the atoms instead of the frontier orbitals in STM measurements[20]. It has been used to visualize the geometric structures of various organic molecules[21-23], graphene[24], non-carbon materials such as water networks on metal surface[25,26]. Recently, 1D water chains on Cu(110) was found to consist of pentagonal water rings by high-resolution nc-AFM imaging[26]. Interestingly, the STM image of 1D water chains on Cu(110) shows zigzagged protrusions resembling those of SiNRs on Ag(110)[27,28].

The STM images and nc-AFM images have been obtained simultaneously from the same sample. Figure 1a shows the STM images of Si SNRs, which features a 2× superstructure with respect to the Ag(110) substrate. The image is consistent with previous reports and our simulated STM image with the pentagon model (Fig. 1b). However, it is beyond the capability of STM to resolve finer structure other than the round, zigzag protrusions along the chain. In contrast, the high resolution nc-AFM image of SNR, as shown in Fig. 1c, gives clearly more details of the structure. Pentagonal rings aligned in a zigzag manner along the chain can be directly visualized, as outlined by red dash lines as a guide to the eyes. We notice that the pentagon structures are slightly distorted, probably due to the inhomogeneous relaxation of the CO molecule at the tip apex during the scanning[29].

It is worth to note that in the pentagon structural model, a row of Ag atoms underneath the Si SNRs is missing. Accordingly, in each Si pentagon, four Si atoms are laying in the missing-silver trough, and the one at the outer corner of the pentagon rests on a silver bridge site (Fig. 1e). This results in a slight buckling of the corner Si atom, with a buckling height of 70 pm[18]. This structural feature perfectly accounts for the relatively bright outer corner of the Si pentagon observed in nc-AFM images. Fig. 1d show the simulated AFM images of the Si SNR with the pentagon model, which fits the experiment very well.

Similarly, the experimental and simulated STM images of a double-strand ribbon (DNR) are shown in Fig. 1f and g. For the DNRs, the nc-AFM image (Fig. 1h) shows many spiral, interconnecting bright lines between two NRs. This feature is perfectly reproduced by the simulated AFM image (Fig. 1i) based on the pentagon model (Fig. 1j), which further supports our structural model. The interconnecting lines should not be automatically related to presence of interatomic bonds, instead they represent ridges of the potential energy landscape experienced by the functionalized probe[30,31].

Raman spectroscopy provides vibrational information of chemical bonds, which serves as chemical fingerprints of molecules. Certainly, pentagon Si rings and hexagon Si rings should present distinctly different spectral fingerprints, which can be another direct evidence of the pentagonal nature of Si NRs. In literature, Raman spectrum of DNRs has been measured with 488 nm excitation laser[32], however, the Raman spectrum of SNRs has not been reported so far. We performed first normal Raman spectroscopy (without tip enhancement) measurements of SNRs and DNRs. In this case, we prepared two different samples, with dominating SNRs and DNRs, respectively, by controlling the sample temperature, as shown in Fig. 2a and (b). It took us a long data accumulation time in order to obtain a reasonable signal-to-noise ratio. The Raman spectra of SNRs and DNRs, as shown in Fig. 2c, are almost identical to each other. This indicates that their atomic structures are the same, with no chemical bonds between the two rows of Si atoms in the middle of the DNRs. There are three major peaks at about 265 cm$^{-1}$, 440 cm$^{-1}$ and 465 cm$^{-1}$, and two weak peaks at 165 and 201 cm$^{-1}$, respectively. These peaks are completely different from the vibrational frequencies base on a hexagonal model of Si NRs[33]. On the other hand, we calculated the vibrational frequencies of the pentagon model of Si SNRs, and the calculated peak positions fits exactly with the experiment ones, as summarized in Table 1. The schematic of the major vibration modes are shown in Fig. 2f. Therefore, our Raman data provides another direct evidence that the Si NRs is composed of pentagonal Si rings instead of hexagonal rings. Possibility of surface alloying can also be ruled out since it will cause significantly different phonon spectrum.

In additional to normal Raman, we further performed gap-mode TERS to clarify the vibrational nature of the Raman modes. Previously, we have utilized tip-enhanced Raman spectroscopy (TERS) to identify different silicene phases which differs only in their Si-Si bond directions on Ag(111) surface[34]. The selective enhancement of Raman modes by TERS provides critical information for identifying the origination of Raman mods. As shown in Fig. 2d, the gap-distance dependent TERS spectra of SNRs shows that the intensity of the $OM_1$ to $OM_4$ peaks, which are below 200 cm$^{-1}$, are strongly enhanced in gap-mode TERS and become the dominant peaks at small tip-sample distance. In contrast, the three major Raman peaks ($IM_1$, $IM_2$ and $IM_3$) get very little enhancement and becomes insignificant. Only the $IM_3$ peak is slightly enhanced and becomes stronger than the $IM_2$ peak. This selective enhancement behavior can be perfectly explained by the vibrational schematics as shown in Fig. 2f. As we know, TERS mainly enhances the vibration modes which contains out-of-plane vibration components, i.e., model with nonzero $\alpha_{zz}$ component in the Raman tensor[35]. In Fig. 2f, one can see that the low energy modes (out-of-plane modes) all have significant vertical components, agreeing with their significant enhancement in TERS. The high energy $IM_3$ also contains clear vertical component, although not as large as the low energy modes. In contrast, the $IM_1$ and $IM_2$ modes have almost purely in-plane vibrations, and therefore they exhibit negligible enhancement in TERS. We shall emphasize that the low frequency modes are too week to perform a reliable analysis in the normal Raman spectra, and it is remarkable that due to the large TERS enhancement they are clearly unveiled and can be reliably analyzed by our model.

In contrast to normal Raman that probes an average signal from a large surface area, TERS is a local probe that provides local Raman spectra with high spatial resolution. We measured the TERS spectra along the line marked in Fig. 3a, with 0.5 nm steps. From the TERS spectra map (Fig. 3b) along the line, one can see the strongest $OM_4$ peak at 202 cm$^{-1}$ vanishes when the tip moves to the bare Ag(110) surface between the two DNR chains, giving a spatial resolution of TERS better than 1 nm. This high spatial resolution guarantees that we can obtain Raman spectrum from very local area underneath the tip, and thus we can compare the TERS spectra of SNRs and DNRs directly. As shown in Fig. 3c, the TERS spectra of SNRs and DNRs were measured in different spots in the same image where SNRs and DNRs coexist on the surface. They were almost identical, consistent with the normal Raman spectra. It further confirms that the Si DNR is two independent rows of Si NRs, without any chemical bonds between them. Notably, the $OM_4$ peaks of the DNRs (202 cm$^{-1}$) show a very slight but clear blue shift from that of the SNRs (200 cm$^{-1}$), as see in the insert in Fig. 3c. The $OM_4$ modes mainly come from the out-of-plane vibration of the top Si atoms ($Si_t$) on silver bridge sites, as see in Fig. 2d. The blue shift is due to the slightly compressed bonds length of $Si_t$-$Si_s$ atoms (about 0.001 Å), for DNRs as compared with SNRs, due to the repulsion interaction between the two Si pentagon chains in DNRs. Note that such weak interaction is observable simply because Raman spectroscopy is extremely sensitive to the structural configuration, and bond length changes as small as 0.001 Å can be detected[36,37]. And due to the strong signal enhancement we are able to obtain a strong signal from the $OM_x$ peaks. Certainly, in the normal Raman $OM_4$ mode of DNRs should also be blue shifted (Fig. 2c), but with a low signal to noise ratio it is impossible to analyze. Similarly, the $IM_2$ stretching (perpendicular to the NR direction) mode is slightly blue shifted and the $IM_3$ stretching (parallel to the NR) mode is slightly red shifted, due to the slightly compressed $Si_s$-$Si_s$ bond in the direction perpendicular the NRs, as shown in Fig. 3d.

Finally, important results were also obtained from Si magic clusters on Ag(110), which co-exist with SNRs at room temperature and low coverage, as has also been reported in previous studies[18,38]. All of the clusters show identical structure in the STM images, with a pair of bright spots in the center and two darker wings (Fig. 4a, f and g). These magic clusters are considered to be the precursor for self-assembly of SNRs. In ref [18] Cerdá et al. proposed a model with a hexagonal Si ring and 4 Si adatoms on a Ag di-vacancy. Based on this assumption, they suggested a transition from hexagon to pentagon structure during the self-assembly of SNRs. Here, we show by nc-AFM and TERS experiments that it is not the case. Firstly, the extremely high spatial resolution of our TERS system allow us to obtain for the first time single-cluster Raman spectrum, as shown in Fig. 4b and 4c. One major Raman peak is observed at 203 cm$^{-1}$ only when the tip is on the top of the cluster, as indicated by the Raman intensity profile of the mode at 203 cm$^{-1}$. This profile was measured along the line in Fig. 4b, in agreement with the topography height profile of the cluster, confirming the extremely high spatial resolution of our TERS system. This 203 cm$^{-1}$ peak is almost identical to the OM$_4$ peak of the SNRs, and differs completely from the calculated results for a hexagonal Si rings with adatoms, whose out-of-plane modes is at 164.5 cm$^{-1}$. In order to explain our data, we proposed a structural model containing two Si pentagons sitting side by side, with four Si adatoms attached to the edge (Fig. 4e). This structure is similar to the pentagon chain model of the SNR, and the calculation indeed gives a main out-of-plane vibrational mode at 207 cm$^{-1}$ (Fig. 4d), consistent with the experiments very well.

To further confirm the structure, high-resolution nc-AFM image is also obtained for the Si magic cluster, as shown in Fig. 4h. The detailed features of the image fitted perfectly the simulated AFM image (Fig. 4i). We note that the pentagon structures cannot be easily distinguished in both the experimental and simulated AFM images as for the cases of SNR and DNR. This is due to the highly buckling nature of the Si cluster, where the central two Si atoms are located deeply in the Ag di-vacancy. In addition, it should be noted that the cluster model is placed on a Ag di-vacancy. Actually, we found that when more Ag atoms underneath the cluster is removed from the surface to form a trough as in the SNR case, the cluster structure becomes unstable, and evolution into the NRs spontaneously. Thus this model can explain the dynamics transition from cluster precursor to NRs as well.

In conclusion, our study combining nc-AFM and Raman spectroscopy provides direct evidences, including the direct visualization and chemical fingerprints, that Si NRs were composed of purely pentagon rings. Furthermore, the structure of the Si magic cluster was found to contain also pentagonal Si rings in contrast to previously proposed hexagonal rings. The dynamics of Si self-assembly process, form clusters to NRs, can be explained by the model as well. We note that Si NRs possesses intriguing properties, such as strong resistance to oxidation[39], and interesting 1D transport characteristics[9,38], hydrogenated Si NRs may possess peculiar magnetic properties such as ferromagnetic or antiferromagnetic[40,41]. These properties based on pentagonal Si rings are not only fundamentally important, but also potentially applicable in nanoelectronic and spintronic applications involving low dimensional silicon structures.

**Acknowledgments**

This work was supported by the National Key R&D Program under Grant No. 2016YFA0300901 and 2017YFA0205003, the National Natural Science Foundation of China under Grant No. 11634001 and 11474277. Y.J. acknowledges support by National Science Fund for Distinguished Young Scholars and Cheung Kong Young Scholar Program.

**Author contributions**

KW designed the experiment. SS, WL and WH performed STM and TERS experiment. RM and DC performed nc-AFM experiment guided by YJ. JW and PT performed Raman calculation and analysis. LK performed first principles structure calculations. All authors contribution to the discussion and analysis of the results.

**Table 1.** Calculated vibrational modes and corresponding frequencies of the pentagon model of Si SNRs.

| Modes | Exp. | Theo. | Diff. |
|---|---|---|---|
| $IM_3$ | 464 | 457 | 7 |
| $IM_2$ | 442 | 440 | 2 |
| $IM_1$ | 265 | 269 | -4 |
| $OM_5$ | 228 | 233 | -5 |
| $OM_4$ | 201 | 208 | -7 |
| $OM_3$ | 163 | 164 | -1 |
| $OM_2$ | 143 | 145 | -2 |
| $OM_1$ | 115 | 110 | 5 |

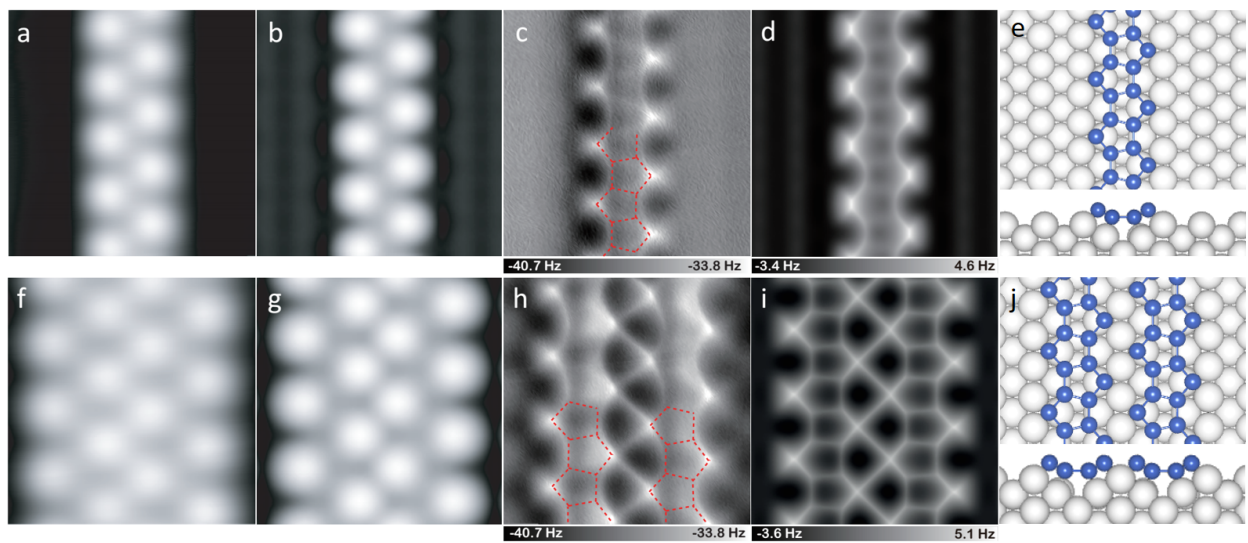

**Figure 1 | Si SNR and DNR on Ag(110) surface. a-d**, STM (**a**) and AFM images (**c**) of Si SNR, and the corresponding simulated ones (**b**, **d**). **e**, Atomic structure of the SNR on Ag(110) in top and side views. Ag, Si atoms are denoted as white and blue spheres respectively. **f-i**, STM (**f**) and AFM (**h**) images of Si DNR, and the corresponding simulated ones (**g**, **i**). **j**, Atomic structure of the DNR on Ag(110) in top and side views. Set point of STM images for (**a**, **f**) are V=100 mV and I=20 pA. The tip height of experimental (simulated) AFM images in (**c** (**d**), **h** (**i**)) are -150 pm (8.91Å) and -200 pm (8.91Å), respectively. The tip height of experimental AFM images is referenced to the STM set point on the Ag surface next to the NRs (100 mV, 50 pA). The oscillation amplitudes of experimental images are 100 pm. All the AFM simulations were done with a quadrupole ($d_{z^2}$) tip (k= 0.5 N/m, Q= 0.0 e).

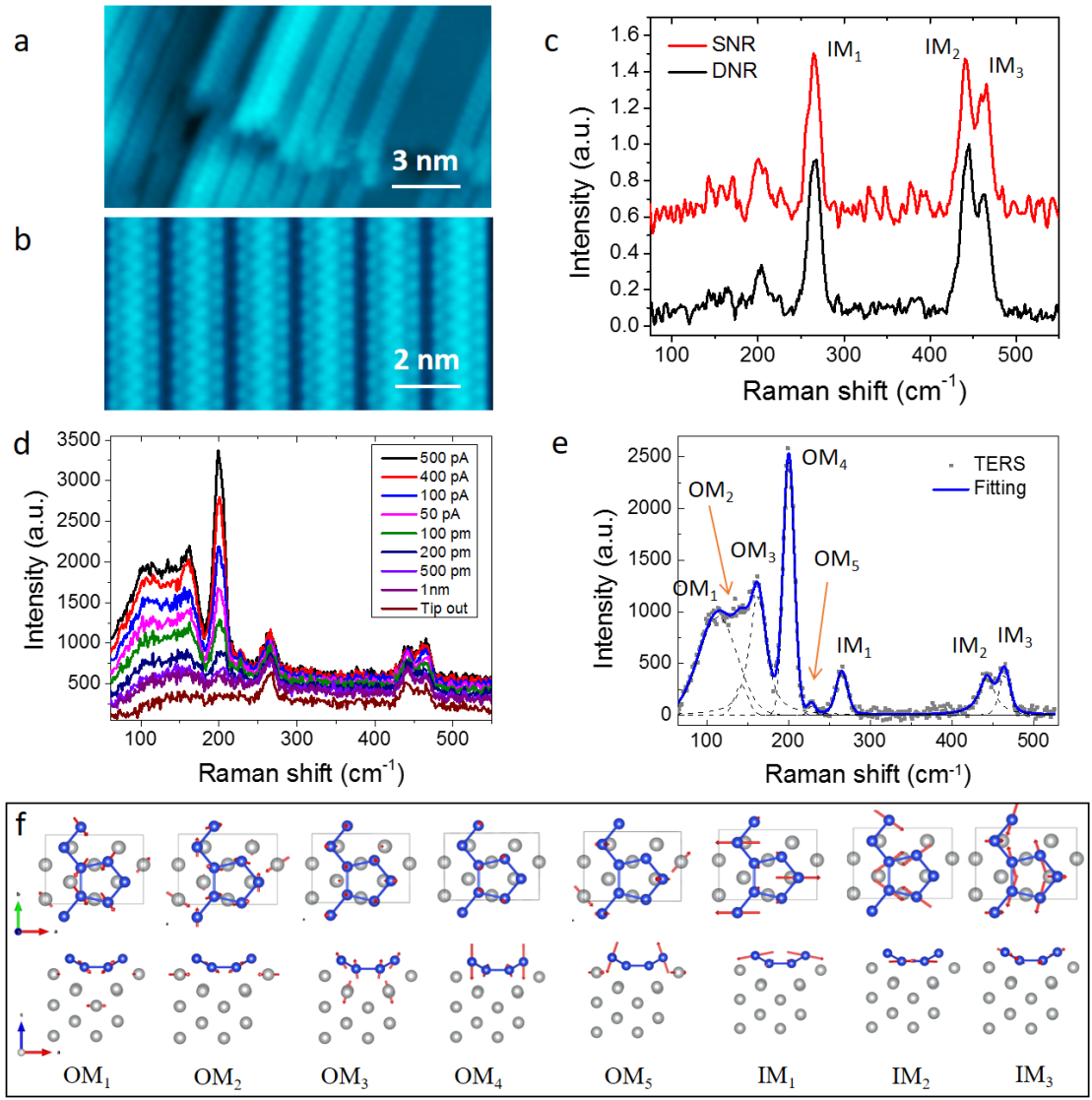

**Figure 2 | Normal Raman and TERS spectra of Si NRs. a**, **b**, STM images of SNRs (**a**) and DNRs (**b**) grown at room temperature and 460 K respectively. **c**, The corresponding normal Raman spectra of SNRs (300 s) and DNRs (600 s) in (**a**) and (**b**). **d**, Gap-distance dependent TERS spectra of SNRs (1 V). **e**, TERS spectrum of SNRs, experimental data are shown by gray crosses, and the fitting data are plotted by blue solid line. **f**, The atomic vibration schematic of each Raman modes.

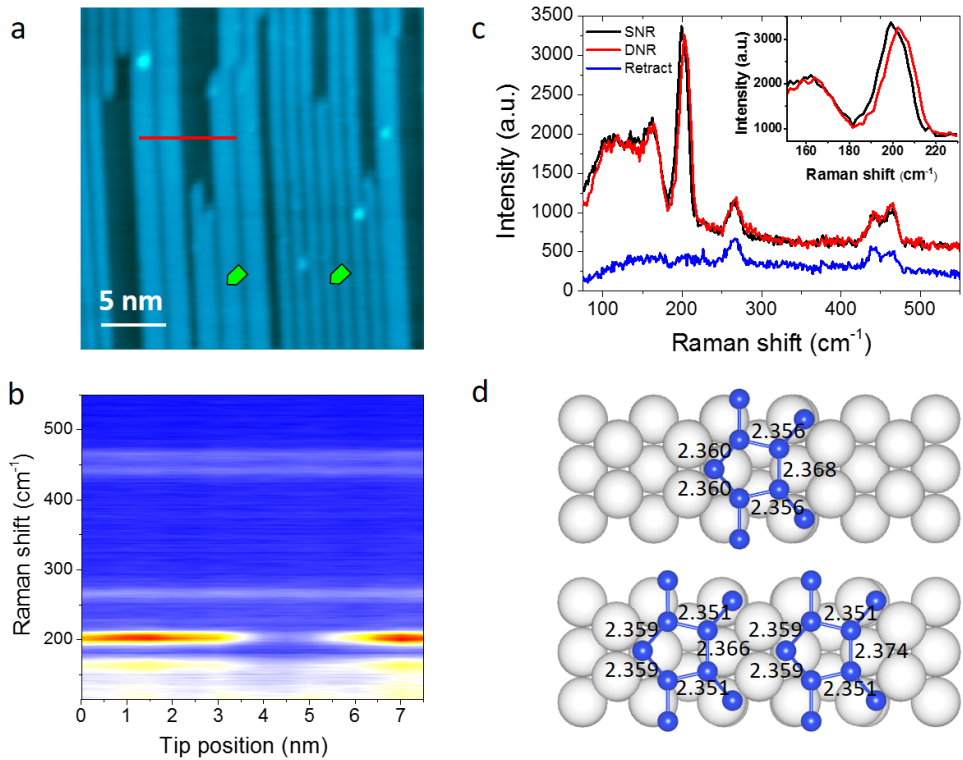

**Figure 3 | TERS spectra of Si SNRs and DNRs. a**, STM topography of Si SNRs and DNRs (1 V, 100 pA). **b**, TERS spectra map along the line in a with an interval of 0.5 nm. **c**, TERS spectra of Si SNRs and DNRs with the tip position labeled in (**a**) (1 V, 100 pA), and Raman spectrum with tip retracted from tunneling. The laser power is about 10 mW, and the acquisition time is 50 s for every spectrum. The inset shows a zoom-in of the main peak. **d**, Calculated Si-Si bonds length of SNR and DNR on Ag(110) in angstrom.

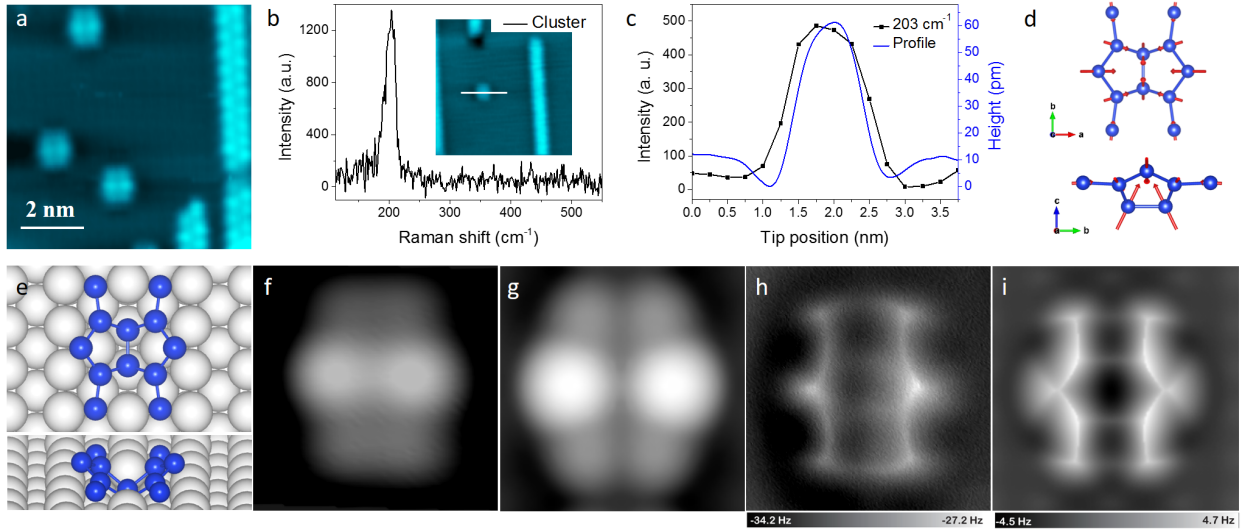

**Figure 4 | TERS spectrum and structure of Si nanocluster. a**, STM of the Si cluster and SNRs at low Si coverage (1 V, 50 pA). **b**, TERS spectrum of the cluster in the inset. **c**, TERS intensity of the 203 cm$^{-1}$ mode along the line in (**b**) with an interval of 0.25 nm every step, and the STM topography height profile. **d**, Atomic vibration schematic of the 203 cm$^{-1}$ mode of the cluster. **e**, Atomic structure of the Si cluster on a Ag di-vacancy Ag(110) surface, with top and perspective views. **f**, High resolution STM images of the cluster (100 mV, 50 pA). **g**, Simulated STM image of the cluster (1 V). **h**, Corresponding AFM image of the cluster in (**f**). The tip height of experimental AFM image is -170 pm, which is referenced to the STM set point on the Ag surface next to the nanocluster (100 mV, 50 pA). **i**, simulated AFM image of the cluster. The tip height in simulations is 9.21 Å, defined as the vertical distance between the apex atom of the metal tip and the highest Ag atom of the substrate. All the oscillation amplitudes of experimental and simulated images are 100 pm. All the AFM simulations were done with a quadrupole (d$_{z^2}$) tip (k= 0.5 N/m, Q= 0.0 e).

## Methods

**Sample preparation.** The clean Ag(110) surface was prepared by repeated cycles of $Ar^+$ sputtering and annealing at 750 K. The Si NRs were grow in-situ in the MBE chamber, with a piece of Si wafer heated to about 1300 K by direct current to deposited Si on the Ag(110) surface. The single-strand Si nanoribbons were obtained with the Ag(110) substrate held at room temperature, and at about 460 K for double-strand Si NRs.

**STM/AFM experiments.** The STM and AFM images in Fig. 1 and Fig. 4 were performed with a combined nc-AFM/STM system (Createc, Germany) at 5 K using a qPlus sensor equipped with a W tip (spring constant $k_0 \approx 1800$ N/m, resonance frequency $f_0 = 26.7$ kHz, and quality factor Q $\approx 60000$). The NaCl(001) bilayer film was grown on the Ag(110) surface to facilitate the preparation of CO-tips. The CO molecules were dosed *in situ* onto the sample surface at 5 K through a dosing tube. The CO-tip was obtained by positioning the tip over a CO molecule on the NaCl film at a set point of 300 mV and 20 pA, followed by increasing the current to 300-400 pA.

**TERS experiments.** The TERS experiment were carried out in a homemade LT-HV-STM based TERS system, at 77 K and with a base pressure about $2 \times 10^{-10}$ mbar. Chemically etched Ag tips were used for STM imaging and TERS measurements, which were degassed at about 800 K in HV for 10 min to remove contaminations from the tip. A p-polarized laser with wavelength at 532 nm was used as pump source for Raman measurement, and the polarization is along the tip axis for TERS experiments. An aspherical lens that can be adjusted by a 3D piezoelectric motor was set inside the low-temperature Dewar for focusing the laser and collecting the scattering Raman light.

**First principle calculation.** Structural relaxation and charge density calculations are performed using the DFT code Vienna ab initio simulation package (VASP)[42] within the projector augmented wave method[43,44] and a plane-wave basis. The exchange correlation potential is treated within the generalized gradient approximation. A 24×8×1 k-mesh is used to sample the BZ for Si NR and an 8×8×1 one for cluster. The energy cutoff for the plane-wave basis is 400 eV. All atoms are fully relaxed until the residual force per atom is smaller than 0.001 eVÅ$^{-1}$. Vibrational frequencies are calculated using DFPT as implemented in VASP[45].

**Simulations of AFM images.** The Δf images were simulated with a molecular mechanics model including the electrostatic force, based on the methods described in refs [46] and [47]. We used the flexible probe-particle tip model: the effective lateral stiffness k = 0.5 N/m and effective atomic radius $R_c$ = 1.661 Å. In order to take the electrostatic forces into account, we added a quadrupole-like charge distribution at the tip apex to simulate the CO-tip for all the AFM simulations as described in ref [48]. The input electrostatic potentials of the Si NRs and clusters were obtained from DFT calculations. Parameters of Lennard Jones (LJ) pairwise potentials for all elements are listed in Table S1.

Table S1. Parameters of Lennard Jones (LJ) pairwise potentials for all elements.

| Element | ε [meV] | r [Å] |
|---|---|---|
| Si | 25.490 | 1.90 |
| Ag | 10.000 | 2.37 |

**Reference**
42. Kresse, G. & Furthmuller, J. Efficient iterative schemes for ab initio total-energy calculations using a plane-wave basis set. *Phys. Rev. B* **54**, 11169-11186, (1996).
43. Kresse, G. & Joubert, D. From ultrasoft pseudopotentials to the projector augmented-wave method. *Phys. Rev. B* **59**, 1758-1775, (1999).
44. Blöchl, P. E. Projector augmented-wave method. *Phys. Rev. B* **50**, 17953-17979, (1994).
45. Baroni, S., de Gironcoli, S., Dal Corso, A. & Giannozzi, P. Phonons and related crystal properties from density-functional perturbation theory. *Rev. Mod. Phys.* **73**, 515-562, (2001).
46. Hapala, P., Temirov, R., Tautz, F. S. & Jelinek, P. Origin of High-Resolution IETS-STM Images of Organic Molecules with Functionalized Tips. *Phys. Rev. Lett.* **113**, 226101, (2014).
47. Hapala, P. *et al.* Mechanism of high-resolution STM/AFM imaging with functionalized tips. *Phys. Rev. B* **90**, 085421, (2014).
48. Peng, J. *et al.* Submolecular-resolution non-invasive imaging of interfacial water with atomic force microscopy. *arXiv:1703.04400v1*, (2017).